\documentclass[runningheads]{svproc}

\usepackage{amsmath,amssymb,amsfonts}
\usepackage{algorithmic}
\usepackage{graphicx}
\usepackage{textcomp}
\usepackage{xcolor}
\usepackage{float}
\usepackage{url}
\usepackage{caption}
\urldef{\mailsa}\path|mukesh.poudel@usm.edu|
\urldef{\mailsb}\path|nick.rahimi@usm.edu|

\usepackage[numbers]{natbib}

\begin{document}

\title{Machine Learning-Based AES Key Recovery via Side-Channel Analysis on the ASCAD Dataset}

\author{Mukesh Poudel\inst{1} \and Nick Rahimi\inst{1}\thanks{Corresponding author}}
\institute{\textsuperscript{1} School of Computing Science and Computer Engineering, University of Southern Mississippi\\Hattiesburg, MS, USA\\
\email{mukesh.poudel@usm.edu, nick.rahimi@usm.edu}}

\maketitle

\begin{abstract}
    Cryptographic algorithms like Advanced Encryption Standard (AES), Rivest–Shamir–Adleman (RSA) are widely used and they are mathematically robust and almost unbreakable but its implementation on physical devices often leak information through side channels, such as electromagnetic (EM) emissions, potentially compromising said theoretically secure algorithms. This paper investigates the application of machine learning (ML) techniques and Deep Learning models to exploit such leakage for partial key recovery. We use the public ASCAD `fixed' and `variable' key dataset, containing 700-sample and 1400 EM traces respectively from an AES-128 implementation on an 8-bit microcontroller. The problem is framed as a 256-class classification task where we target the output of the first-round S-box operation, which is dependent on a single key byte. We then evaluate standard classifiers (Random Forest (RF), Support Vector Machine (SVM)), a tailored Convolutional Neural Network (CNN) and a Residual Neural Network(ResNet). We also explore the utility of RF-based feature importance for dimensionality reduction. Crucially, we employ this domain-specific Key Rank metric for evaluation, showing its necessity over standard classification accuracy, which remained below 2\% due to low signal-to-noise ratio. Our results show that SVM and RF on full features perform poorly in key ranking. However, RF trained on reduced (top 100) identified via importance analysis achieves Rank 0 (successful key byte recovery) using almost half the attack traces. The implemented CNN as well, despite exhibiting overfitting in terms of validation loss, also achieves Rank 0 efficiently using approximately 65 attack traces for the fixed-key dataset. The ResNets perform best on large and complex datasets but may not always be the best choice for simple fixed key dataset in terms of efficiency. Thus we conclude that models, particularly CNNs, ResNets and feature-selected RF, coupled with the Key Rank metric, are an effective tool for side-channel key recovery, confirming the practical vulnerability of the cryptographic implementations.
\end{abstract}


\section{Introduction}
Cryptographic algorithms like the Advanced Encryption Standard (AES) are mathematically robust and cannot be compromised through mathematical flaws. However, physical devices executing cryptographic operations lead to unintentional information leakage through various `side channels', such as power consumption, timing variations, and electromagnetic (EM) emissions.\cite{obaid2025enhancing} Electromagnetic analysis (EMA) is a potent form of side-channel analysis (SCA) where attackers non-invasively measure the EM fields radiating from a device during any cryptographic operation. These emissions often contain subtle variations correlated with the intermediate data being processed. And these are often related to the key, which poses a huge challenge. Recent advancements have shown that Machine Learning (ML) and Deep Learning (DL) are powerful tools for automatically learning these complex correlations, potentially outperforming traditional statistical SCA techniques.\cite{berreby2023investigating}\cite{kocher1999differential}

This paper investigates the practical application of ML techniques to recover an AES-128 key byte by analyzing EM side-channel traces from the public ASCAD (ANSSI SCA Database) fixed and variable key dataset\cite{benadjila2020deep}. We frame the problem as a multi-class classification task targeting the output of the first-round AES S-box operation. The low signal-to-noise ratio in the EM traces poses a significant challenge for standard classification metrics like accuracy. This often yields misleadingly poor results. Therefore, a key aspect of this work is the rigorous use of the domain-specific Key Rank metric. Key Rank evaluates an attack's success by determining the position of the true key in a list of all possible key candidates ranked by their likelihood score derived from the ML model's output across multiple traces. This metric directly reflects the practical ability to recover the key, even when per-trace classification accuracy is low.

The contributions of this work are: (1) a comparative performance analysis of standard classifiers (Random Forest (RF), Support Vector Machine (SVM)), ResNets and a tailored Convolutional Neural Network (CNN) for AES key byte recovery on the ASCAD dataset, (2) an exploration of RF-based feature importance for dimensionality reduction, its impact on model efficiency and effectiveness, (3) a clear demonstration of the necessity of the Key Rank metric over the standard accuracy for evaluating ML-based SCA success, and (4) confirmation of successful key recovery using ResNets, CNN, SVMs and feature-selected RF models, despite low classification accuracy, highlighting the practical feasibility of ML-based side-channel attacks.

The remainder of this paper is organized as follows: Section 2 details the AES S-box operation, EM leakage principles, the ASCAD dataset, and outlines our experimental setup and the Key Rank evaluation methodology. Section 3 presents and analyzes the results from the RF, SVM, CNN and ResNet models. Section 4 discusses the implications of our findings, the accuracy versus Key Rank paradox, and limitations. Finally, Section 5 concludes the paper and suggests avenues for future research.

\section{Background and Methodology}
This section provides the necessary background on the target cryptographic operation, the nature of EM side channel leakage, the ASCAD dataset used for experiments, and the specific ML models implemented in our study. We begin with an overview of the AES S-box operation, followed by a discussion of the ASCAD dataset and the data preprocessing steps. We then detail the architectures and hyperparameters for our implemented ML models (Random Forest, SVM, CNN and ResNets) and finally reiterate the importance of the Key Rank metric for evaluation. 
\subsection{AES - Sbox Operation and Leakage }
The Advanced Encryption Standard (AES) is a symmetric block cipher that encrypts data in 16-byte (128-bit) blocks using a key of 128, 192, or 256 bits. AES operates in rounds, with the number of rounds depending on the key size (10 rounds for AES-128, 12 for AES-192, and 14 for AES-256). Each round involves several transformations, including SubBytes, ShiftRows, MixColumns, and AddRoundKey.\cite{huang2025deep} A fundamental non-linear operation within the SubBytes step is the application of the AES S-box substitution, independently to each byte of the internal state. The S-box input for a given byte position $i$ is the result of $\text{Plaintext}[i] \oplus \text{Key}[i]$, where $\oplus$ represents the XOR operation. The S-box then outputs a different byte based on its lookup table:

\begin{equation}
\textit{Sbox\_Output}[i] = \textit{Sbox} (\textit{Plaintext}[i] \oplus \textit{Key}[i])
\end{equation}

\begin{figure}[htbp]
    \centering
    \includegraphics[width=0.3\textwidth,height=0.3\textheight]{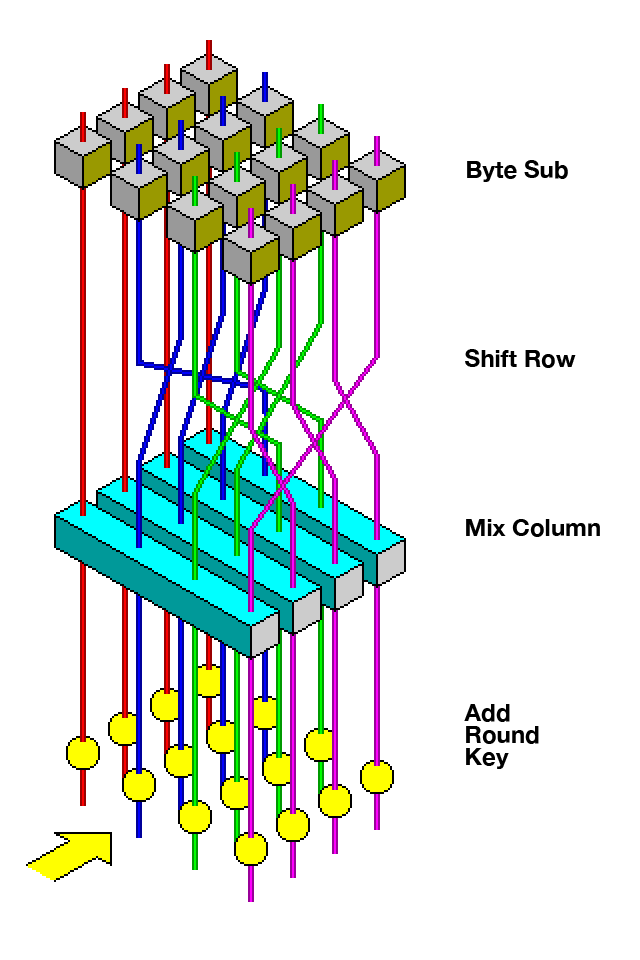}
    \caption{Basic Steps of AES Encryption Round (source: \cite{aes_round_commons})}
    \label{fig:aes_steps}
\end{figure}

Figure~\ref{fig:aes_steps} illustrates the four basic steps within each round of AES encryption (excluding the final round which omits MixColumns).

The S-box operation represents a critical point of vulnerability in an AES implementation. Mangard and Schramm identified that although linear operations at the beginning of the S-box do not leak any information, significant leakage occurs in the first masked multiplier where `the XOR gates of this multiplier absorb a different number of transitions for different data inputs.'\cite{mangard2006pinpointing} These differential transitions create a distinctive power consumption pattern that correlates directly with the processed data values and creates the side-channel leakage we aim to exploit. Since the input to the S-box in early rounds (like the first round we target) directly combines known plaintext with the unknown key byte, any leakage related to this operation provides information about the key. 

We adopt the common value-based leakage model, assuming the EM trace contains information correlated with the specific identity (value 0-255) of the $\text{Sbox\_Output}[i]$. Since this output byte depends on both the known $\text{Plaintext}[i]$ and the unknown $\text{Key}[i]$, predicting the S-box output from the trace allows us to deduce $\text{Key}[i]$. As an 8-bit byte can take 256 distinct values, predicting the S-box output becomes a 256-class classification problem. We specifically target the 3rd byte (index $i=2$) in this study.

\subsection{ASCAD Datasets : Fixed-Key (ASCADf) and Variable-Key (ASCADv)}
For this project, we utilize the publicly available ASCAD `fixed-key' dataset, provided by ANSSI.\cite{benadjila2020deep} To further improve the robustness of the algorithm and to mimic a more realistic scenario, we also trained on the 'variable-key' dataset. This dataset presents a more challenging scenario where the secret key changes for each trace in both the profiling and attack sets. We will be referring to the fixed-key dataset as ASCADf and the variable-key dataset as ASCADv throughout this paper for simpler referencing. The ASCAD datasets were chosen for this research due to it being publicly available and well-documented. This is a widely used benchmark in the SCA community and thus helps in reproducible research and comparison amongst existing literature. 

This particular dataset comprises EM measurements from an AES-128 implementation on an 8-bit ATMega8515 microcontroller, where fixed 128-bit key (for ASCADf) or variable 128-bit keys (for ASCADv) were used. For a supervised Machine Learning problem, this dataset includes precisely labeled data (S-box output for a known key byte ) and corresponding plaintext to help train the classification model. The datasets contains:

\begin{itemize}
    \item \textbf{Profiling Traces:} For ASCADf, A set of 50,000 traces used for training models. Each trace consists of 700 EM measurements (features). For ASCADv, A set of 200,000 traces were used for training and each trace consists of 1400 EM measurements.  Associated metadata includes the plaintext, the fixed key, and pre-calculated labels representing the output of the S-box for the 3rd key byte (index 2).
    \item \textbf{Attack Traces:} A set of 10,000 traces (ASCADf) and 100,000 traces (ASCADv) were used for testing and evaluating the trained models. These traces also contained same number of EM measurements as the profiling counterparts, along with corresponding plaintexts. During the attack phase, the key is treated as unknown.
\end{itemize}

\subsection{Data Preprocessing}
The raw EM traces have different offsets and scales which can impact the performance of Machine Learning models. To fix this, we standardize the training and testing data for both ASCADf (700 traces) and ASCADv (1400 traces) using the mean and standard deviation calculated solely from their respective profiling sets (50,000 traces for ASCADf, and 200,000 traces for ASCADv profiling). This helps make sure that each feature has a zero mean and unit variance. Without scaling, features with higher magnitudes can disproportionately affect the model's decisions, leading to suboptimal performance. Many ML algorithms, particularly those using gradient descent (like CNNs) or distance measures (like SVMs with RBF kernels), perform better and converge faster when input features are on a similar scale and centered around zero. We use Scikit-learn's \textit{StandardScaler} to perform this standardization. We fit the scaler on the training data and then transform both the training and testing data using the fitted scaler. This ensures that the test data is transformed in the same way as the training data, preventing any information leakage from the test set into the model training process.

\subsection{Machine Learning Models}

\textbf{Random Forest (RF):} It is an ensemble learning method that aggregates the predictions of multiple decision trees. We employ Scikit-learn's RandomForestClassifier with $n\_estimators=100$. This provides a good balance between performance and computation. To mitigate overfitting on the high-dimensional data, we apply regularization by setting $max\_depth=20$ to limit the tree complexity and $min\_samples\_leaf=10$ to ensure that each leaf node has sufficient number of samples. Furthermore, RF provides Gini importance scores, which we leverage for feature selection. We sorted features by their Gini importance scores in descending order and selected the top 100 most important features. We train one RF model using all 700 features for ASCADf and 1400 features for ASCADv and another using only the top 100 features as determined by the importance ranking.

\textbf{Support Vector Machine (SVC):} It is a classifier that aims to find an optimal hyperplane to separate different classes. Due to its high computational cost and as it scales poorly with the addition in number of samples and features, we train Scikit-learn's SVC only on the reduced set of 100 features selected by RF. We use the default RBF kernel \textit{(C=1.0, gamma='scale')} and enable probability estimates i.e \textit{probability=True} as it is required for key ranking which relies on the class probability estimates.

\textbf{Convolutional Neural Network (CNN):} We employ a custom CNN architecture implemented in PyTorch, drawing inspiration from ASCAD paper in deep learning-based SCA. The network is designed to learn relevant features directly from the raw EM traces, enabling effective classification of S-box outputs. The architecture details are as follows:
\begin{figure}[hbtp]
    \centering
    \includegraphics[width=1\textwidth]{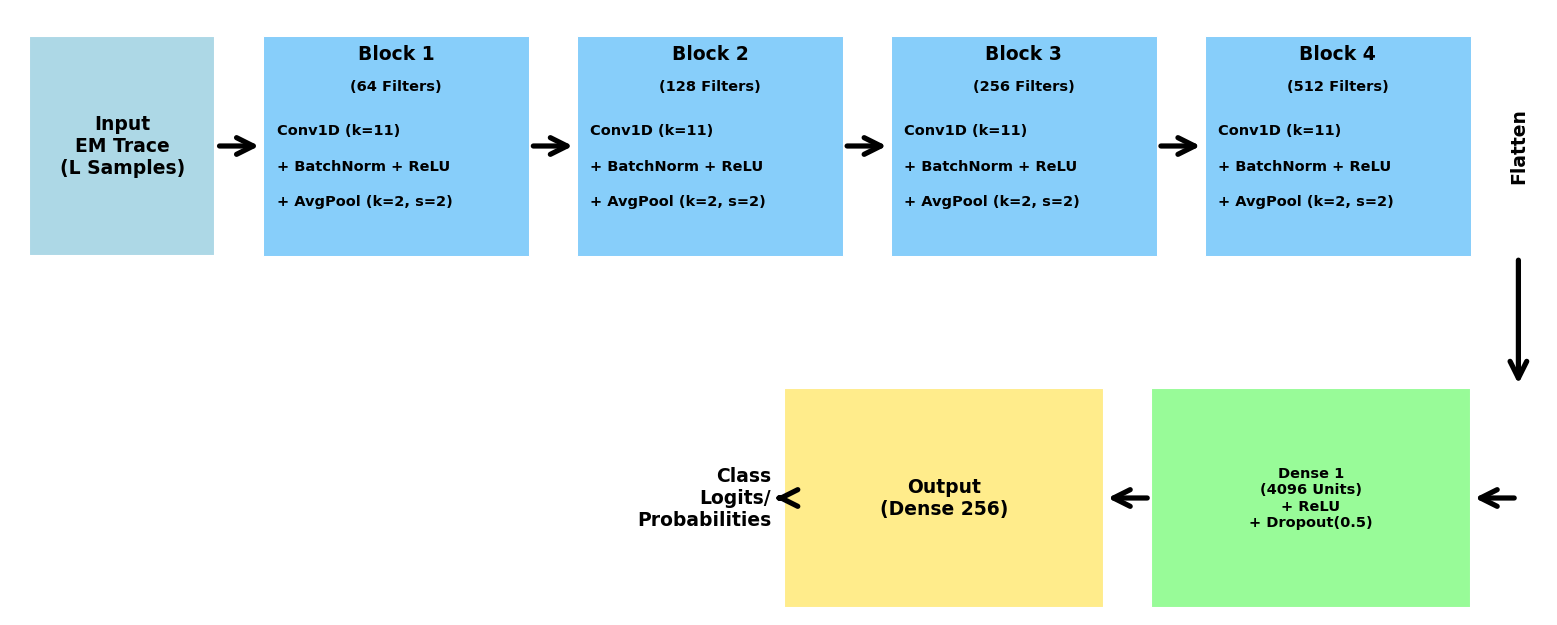}
    \caption{CNN Architecture for SCA}
    \label{fig:cnn_architecture}
\end{figure}

\begin{itemize}
    \item \textbf{Input Layer:} Accepts a 1D EM trace (\textbf{700 samples} for ASCADf, \textbf{1400 samples} for ASCADv) as a tensor of shape \texttt{(BatchSize, 1, TraceLength)}
    \item \textbf{Convolutional Blocks (4x):} The network employs four identical convolutional blocks for hierarchical feature extraction. Each block includes:
        \begin{enumerate}
            \item \textit{Conv1d Layer:} Applies 1D convolutions with a kernel size of 11 and padding of 5 (effectively 'same' padding for the convolution operation itself). This choice is directly supported by findings in \cite{benadjila2020deep} which demonstrated that a larger kernel size (e.g., 11) significantly improves SCA-efficiency, especially compared to multiple layers with smaller kernels. The number of output channels (filters) increases progressively: 64, 128, 256, and 512 for the four blocks, respectively, increasing feature map depth to compensate for spatial dimension reduction by pooling layers, allowing the network to learn a richer set of features.
            \item \textit{BatchNorm1d Layer:} Performs batch normalization.
            \item \textit{ReLU Activation:} Applies the Rectified Linear Unit activation function.
            \item \textit{AvgPool1d Layer:} Performs average pooling with a kernel size of 2 and a stride of 2, downsampling the feature map length by a factor of 2 at each block.
        \end{enumerate}
    \item \textbf{Flatten Layer:} Converts the 2D feature maps (channels $\times$ length) from the final pooling layer into a 1D vector. After four pooling layers, an initial trace of length $L$ becomes $L/16$ (with floor operations for odd lengths at intermediate stages).
    \item \textbf{Dense Head (Classification):}
        \begin{enumerate}
            \item \textit{Linear Layer (4096 units):} A fully connected layer with 4096 output units, followed by ReLU activation. The input size to this layer is dependent on the initial trace length:
            \begin{itemize}
                \item For ASCADf (700 input samples, length becomes 43 after pooling): $512 \text{ channels} \times 43 \text{ features} = 22016 \text{ inputs}$.
                \item For ASCADv (1400 input samples, length becomes 87 after pooling): $512 \text{ channels} \times 87 \text{ features} = 44544 \text{ inputs}$.
            \end{itemize}
            \item \textit{Dropout Layer (p=0.5):} Applies dropout regularization.
            \item \textit{Linear Layer (256 units):} A final fully connected layer outputting logits for the 256 possible S-box values.
        \end{enumerate}
    \item \textbf{Training Parameters:}
        \textit{Optimizer} : RMSprop optimizer with a learning rate of $1 \times 10^{-5}$ and a weight decay of $1 \times 10^{-5}$, \textit{Loss Function} : CrossEntropyLoss, \textit{Batch Size} : 100 \textit{Epochs} : 150.
\end{itemize}
The specific CNN architecture is inspired by successful models described in recent literature, and the original ASCAD paper, but has been modified to suit our computational resources and dataset characteristics.\\

\textbf{Residual Neural Network (ResNet):}
To investigate the performance limitations of the baseline CNN architecture on the more complex variable-key dataset, we implemented a more advanced model based on a Residual Network (ResNet). This architecture is specifically designed to ease the training of deeper networks by introducing \textbf{residual blocks} that help mitigate the vanishing gradient problem. \cite{he2016deep}The overall structure consists of a feature extractor built from a series of these residual blocks, followed by the same dense classification head used in the standard CNN.

\begin{figure}[hbtp]
    \centering
    \includegraphics[width=1\textwidth]{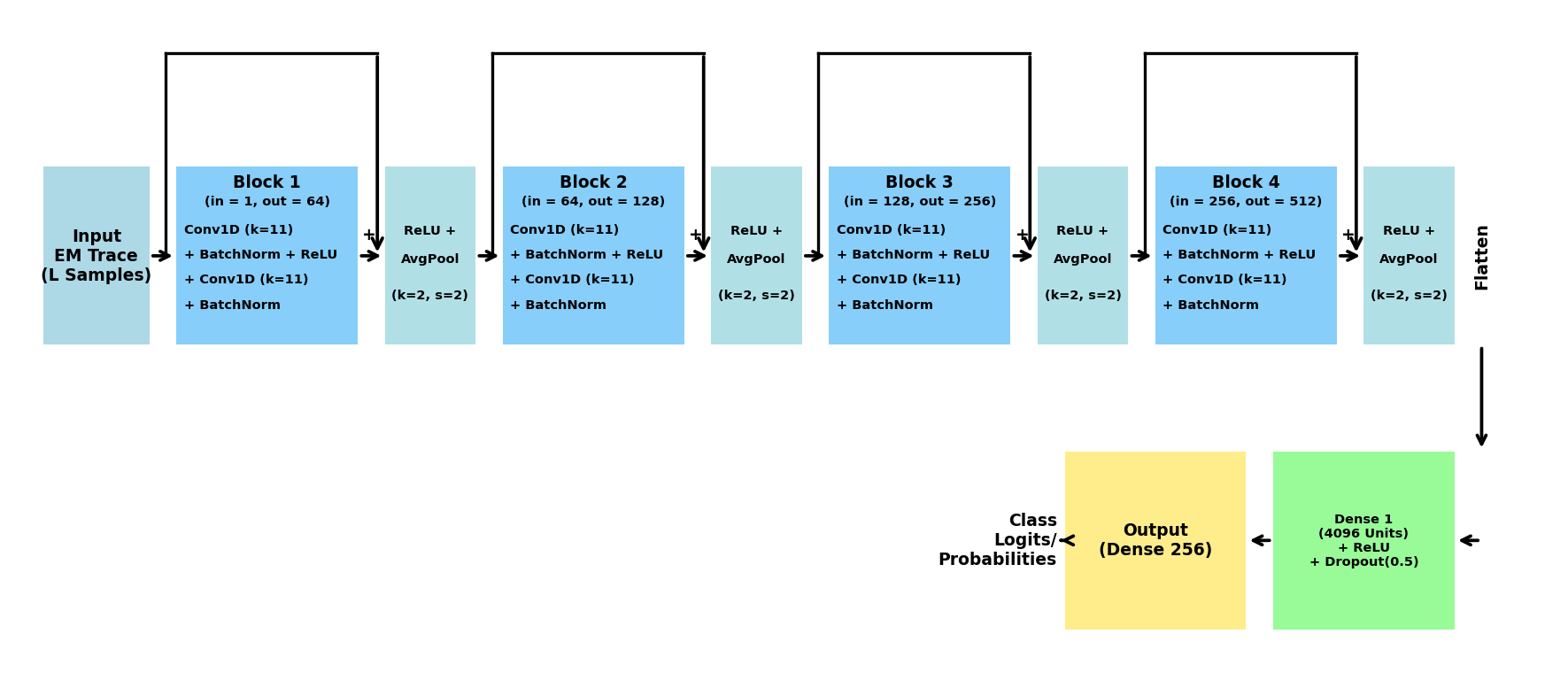}
    \caption{ResNet Architecture for SCA}
    \label{fig:resnet_architecture}
\end{figure}
The core of this model is the Residual Block, which processes the input through two parallel paths:
\begin{itemize}
    \item \textbf{The Main Path:} This path consists of two 1D convolutional layers, each with a kernel size of 11. Each convolution is followed by a Batch Normalization layer, and a ReLU activation function is applied after the first block.
    \item \textbf{The Shortcut Connection:} This path bypasses the main convolutional layers, allowing information to flow directly to a later stage of the block. If the number of input channels does not match the number of output channels for the block, a 1x1 convolution is applied in the shortcut path to match the dimensions. Otherwise, it acts as an identity connection.
\end{itemize}

The outputs of the main path and the shortcut connection are then summed element-wise. A final ReLU activation is applied to this sum, and the result is passed to the next layer.

Our full ResNet feature extractor is composed of four of these $ResidualBlocks$ stacked sequentially. Following each residual block, an $AvgPool1d$ layer with a kernel size of 2 and a stride of 2 is used to downsample the feature map length by a factor of two. Similar to the baseline CNN, the number of filters in the convolutional layers increases through the network, using values of 64, 128, 256, and 512 for the four blocks, respectively. After the feature extractor, the data is flattened and passed to a dense classification head identical to the one described for the standard CNN.
\subsection{Hyperparameter Selection}
To get a deeper understanding and analysis of our model configurations and justify our parameter choices, we conducted a series of runs with varied hyperparameters, which helped validate our baseline parameters and provide insights into the model's sensitivities. \\

\textbf{Random Forest}: Our baseline RF model with 100 features proved highly effective and computationally efficient. To justify our parameter choices, we explored variations. We found that increasing the number of trees $n\_estimators=200$ offered no improvement in attack efficiency (180 traces) while increasing the computation time. Furthermore, using a more complex model by increasing the tree depth ($max\_depth=30$) caused the model to fail in recovering the key, highlighting the importance of a constrained depth for regularization against noisy data. These findings validate that our baseline parameters represent an effective balance between performance and cost.

\textbf{Support vector Machines(SVC)}: The baseline SVC was highly effective, recovering the key in 320 traces. While successful, the SVC was the most computationally expensive model, with training being particularly demanding for the larger variable-key dataset. To justify our parameter choices, we focused on tuning the $c$ parameter, which controls the model's regularization. The $c$ parameter manages the trade-off between creating a simple model (with a wide margin) and correctly classifying all training points (a more complex model). A $c$ value that is too high can cause the model to overfit by memorizing noise in the training data. An overfitted model, by contrast, may make highly confident but incorrect predictions on the attack set, which severely penalizes the cumulative log-probability score of the true key. Our exploration confirmed that $c$=1.0 provided a strong, generalizable result, while a higher value of $c$=10.0 or $c$=100.0 did not yield a significant performance gain to justify the increased risk of overfitting and the even longer training times. Thus we concluded that the standard $c$=1.0 and gamma='scale' parameters provided strong results. 

\textbf{CNN}: Our primary CNN architecture, as described in Section 2.4, is highly effective but sensitive to its training configuration. The baseline model with $kernel\_size=11$, and $batch\_size=100$ consistently recovered the key on the fixed-key dataset in approximately 65 traces. Performance was particularly sensitive to batch size. Decreasing the batch size to 64 required 90 traces, but resulted in longer training times. Similarly, while increasing the batch size to 200, the training time was reduced, but it required almost 180 traces. Further exploration were made by varying the kernel size and the network depth. But it did not yield a better result than our baseline choice of 11. We experimented with 3 blocks and 5 block architecture, but 3 block were not sufficient to recover key, and 5 block architecture was a tradeoff on time. Critically, this tuning process also confirmed the CNN's limitations, as all tested configurations failed to recover the key on the more complex variable-key dataset. This current architecture does not generalize well to the ASCADv challenge, which suggests that more advanced network designs are required for such complex scenarios. This problem is addressed by ResNets.

\textbf{ResNet}: To address the generalization failure of our standard CNN on the ASCADv dataset, we introduced a ResNet architecture. Our hypothesis was that the plain CNN, despite its depth, suffered from degrading gradient flow that prevented learning the abstract patterns required for variable-key attacks. The ResNet architecture directly addresses this through residual blocks with shortcut connections that add the block's input to its convolutional output, We tested the ResNet using a configuration directly comparable to our best standard CNN: four blocks, $kernel\_size=11$, and similar filter progression. ResNet succeeded exceptionally well where the standard CNN had failed, confirming that residual connections were the critical factor for learning generalizable features. While we explored minor variations in depth and kernel size, our chosen configuration represented an optimal trade-off between performance and complexity, solidifying our conclusion that the architectural shift was the principal reason for success.

\subsection{Evaluation Metric: Key Rank}
While standard classification accuracy was measured, it proved uninformative due to high noise levels. The primary metric for SCA success is the Key Rank.\cite{picek2019curse} The process is as follows:

\begin{itemize}
    \item Obtain the predicted probability distribution (vector of 256 probabilities) from the trained model for each of the N attack traces. Let P (label=z\textbar trace\_i) be the probability assigned to S-box output value z for trace i.

     \item For each key byte hypothesis k\_guess (from 0 to 255):
        \begin{itemize}
            \item Calculate the hypothetical S-box output \\$\textit{Z\_hyp\_i} = \textit{Sbox}(\textit{$\text{plaintext\_i}$} \oplus \textit {$\text{k\_guess}$})$ for each attack trace $i$ from 1 to $N$.
            \item Calculate the total Summed Log-Probability (SLP) score for this key guess:\\
            $\textit{Score} (\text{k\_guess})=\sum_{i=1}^{N} \log (P (\textit{label}=Z\_hyp\_i | \textit{trace}_i) + \varepsilon)$ \\
            where $\varepsilon$ is a small constant (e.g., 1e-40) to prevent $\log(0)$. The logarithm is used to avoid numerical underflow when multiplying many small probabilities, and to improve computational efficiency by converting multiplications to additions.
        \end{itemize}
    \item Rank the 256 key guesses based on their scores, from highest (most likely) to lowest.

    \item The Rank of the true key byte (k\_true) is its position in this ranked list (Rank 0 indicates it has the highest score and is successfully recovered).

    \item Plotting the Rank vs.\ the number of traces (N) shows the efficiency of the attack. The Key Score Plot (bar chart of Score(k\_guess) vs.\ k\_guess) visually confirms the correct key's dominance.
    
\begin{figure}[htbp]
    \centering
    \includegraphics[width=0.5\textwidth]{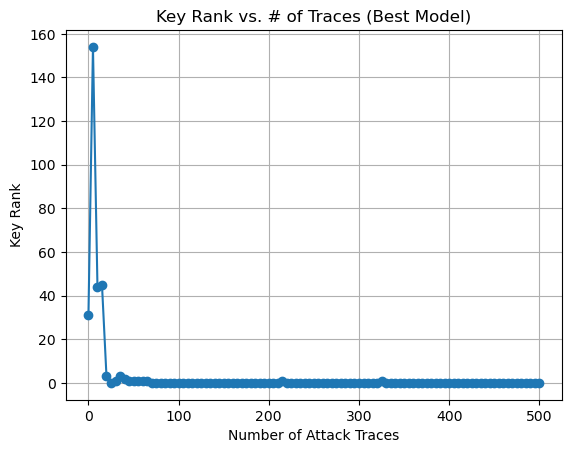}
    \caption{Sample Example of a Key Rank Chart}
    \label{fig:key_rank_example}
\end{figure}
Figure \ref{fig:key_rank_example} illustrates a sample Key Rank chart, showing the rank of the true key byte (2) as a function of the number of attack traces used. The graph illustrates how the rank improves with more traces, and successfully recovers the key after ~50 traces.
\end{itemize}
The superiority of Key Rank over standard accuracy in this context stems from the nature of SCA. The goal is not to achieve perfect classification of the S-box output for every single noisy trace, which is often an unrealistic expectation due to low signal-to-noise ratio (SNR). Instead, the goal is to distinguish the single correct secret key byte from 255 incorrect hypotheses. Key Rank achieves this by aggregating subtle, consistent evidence (the model's probability assignments) across numerous traces. Even if a model has low accuracy, if it consistently assigns a slightly higher probability to the true S-box output (when the correct key is hypothesized) compared to random outputs, this difference gets amplified when log-probabilities are summed over many traces. This makes the score for the true key dominant, leading to successful recovery, while individual trace noise that confounds accuracy is averaged out.
\section{Experimental Results}
\subsection{Setup}
Experiments were conducted using Python 3.11 with PyTorch (for CNN and ResNet) and Scikit-learn (for RF, SVM, Scaler) libraries. The Random Forest model utilized 16 cores for parallel processing and 16GB of RAM, while the SVM was limited to a single thread. Training and evaluation were performed on an Nvidia P100 GPU for the CNN and ResNet models and CPU for RF and SVM models.

\subsection{Random Forest Results}
For the full-feature RF model on ASCADf ($n\_estimators=100$, $max\_depth=20$,$min\_samples\_leaf=10$), 10-fold cross-validation yielded a training accuracy of 43.55\% but a validation accuracy of only 0.46\%. This highlights the challenge of directly classifying individual traces due to the low signal-to-noise ratio inherent in EM leakage. The Key Rank metric is therefore crucial for evaluating SCA success.

In terms of cross-validation, 5 out of 10 folds achieved Rank 0 within 1000 traces. Among these successful folds, the average number of traces required was 492 ($\sigma = 129.21$). The final full-feature RF model exhibited similar performance, requiring hundreds of attack traces to recover the key.

Feature selection using RF's Gini importance significantly improved performance. By training a second RF model on only the top 100 features, the number of attack traces required to achieve Rank 0 was reduced to approximately 200. This represents a 50\% reduction in the number of traces needed for successful key recovery compared to the full-feature model, demonstrating the effectiveness of dimensionality reduction in mitigating overfitting and focusing on the most informative leakage points.

RF on ASCADv with similar parameters and full 1400 features required 750 traces to recover the key. Meanwhile, evaluation on reduced 100 features recovered the key in only 470 traces, reducing the trace count by almost 40\% and re-emphasizing the importance of feature-reduction. 

\begin{figure}[H]
    \centering
    \begin{minipage}[b]{0.48\textwidth}
        \centering
        \includegraphics[width=\textwidth]{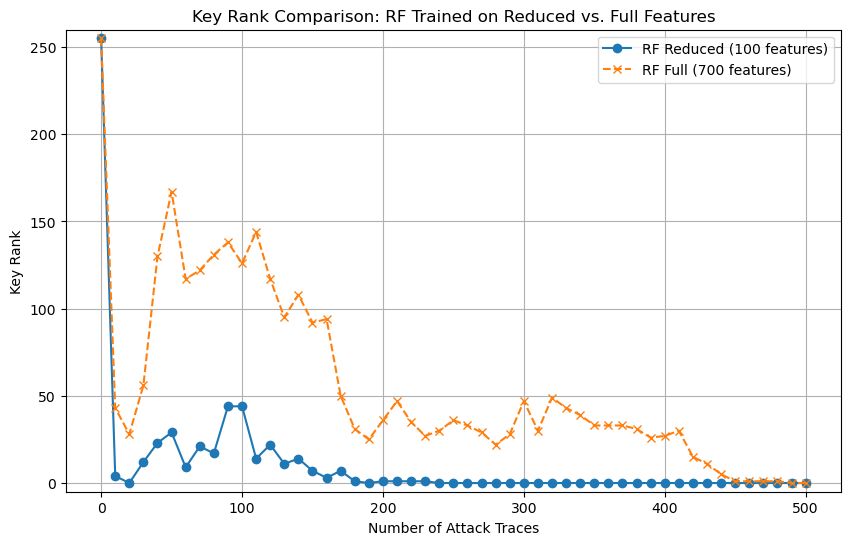}
    \end{minipage}
    \hfill
    \begin{minipage}[b]{0.48\textwidth}
        \centering
        \includegraphics[width=\textwidth]{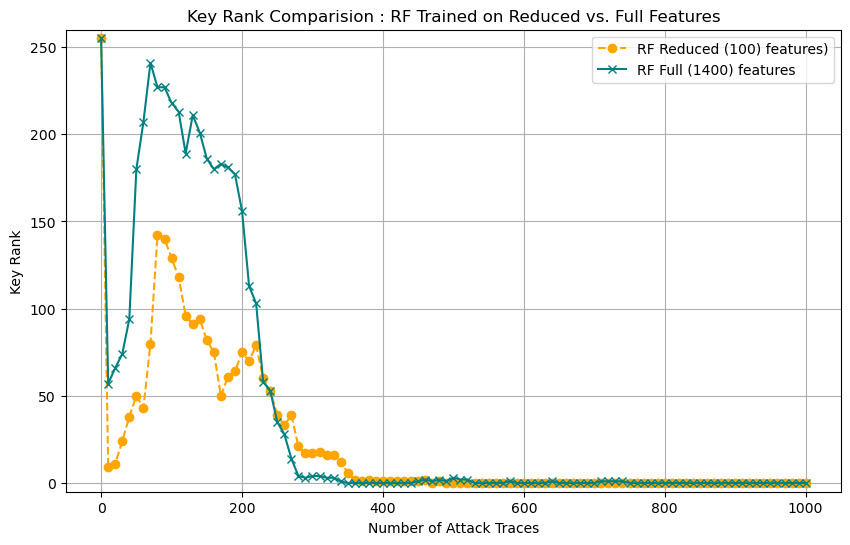}
    \end{minipage}
    \caption{Key Rank Charts for RF on Datasets ASCADf(left) and ASCADv(right)}
    \label{fig:rf_key_rank_combined}
\end{figure}

The feature importance analysis also revealed that leakage is distributed across the trace but with clear concentration in certain time regions. This confirmed the targeted S-box operation leaves electromagnetic fingerprints at specific points in time during execution.

\subsection{CNN Results}
The CNN model exhibited typical deep learning behavior with continuously decreasing training loss (from 5.56 to 5.27) but relatively stable validation loss (around 5.38-5.40), suggesting overfitting by conventional metrics. The final test accuracy was extremely low at only 0.81\%, which would typically indicate a failed model.
\begin{figure}[htbp]
    \centering
    \begin{minipage}[b]{0.48\textwidth}
        \centering
        \includegraphics[width=\textwidth]{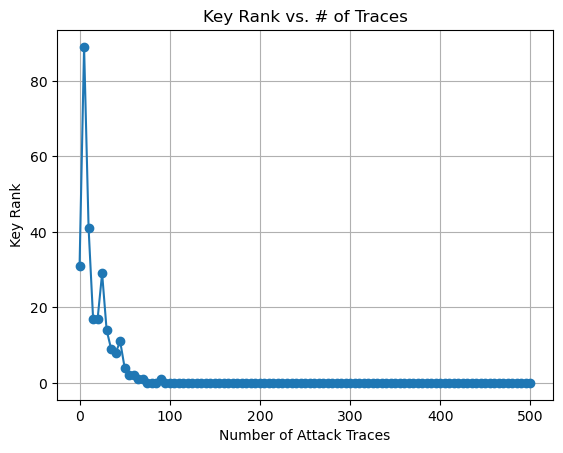}
    \end{minipage}
    \hfill
    \begin{minipage}[b]{0.48\textwidth}
                \centering
                \scalebox{1}[1.22]{\includegraphics[width=\textwidth]{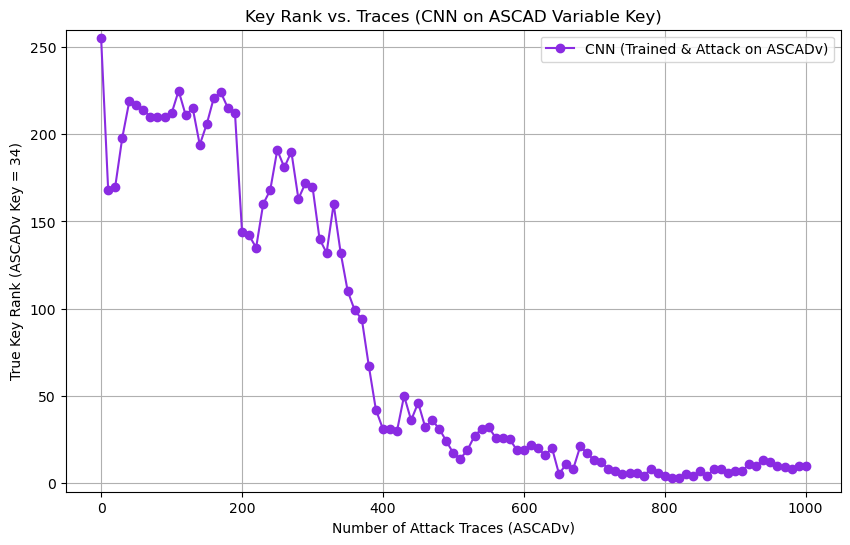}}
        \end{minipage}
    \caption{Key Rank Charts for CNN on ASCADf(left) and ASCADv(right)}
    \label{fig:rf_key_rank_combined}
\end{figure}

However, the key recovery performance told a completely different story. When tested on the attack set, the CNN model's rank of the correct key byte dropped rapidly, reaching Rank 0 consistently after approximately 65 attack traces. This performance was superior to both the full-feature and reduced-feature RF models and is consistent with recent literatures. A reduced learning rate and reduced batch size consistently improved performance. We decided that the current batch size of 100 and learning rate of $1 \times 10^{-5}$ were best in terms of performance and computational efficiency. 
The final key score distribution after using all attack traces showed a clear, dominant peak at the correct key byte 224(for ASCADf), far exceeding the scores of incorrect key guesses. This confirms the CNN successfully learned relevant leakage patterns despite its low classification accuracy. The CNN trained on ASCADv told a different story. The key-byte recovery was unsuccessful with unstable potentially recovery at 800 traces but it was not sufficient enough. This tells us that there is room for improvement with potential architectural changes for the ASCADv dataset.  

\subsection{Support Vector Machine Results}
The SVM model was trained only on the reduced set of 100 features selected by RF feature importance, as training on the full 700 and 1400 features would be computationally complex and time consuming. Despite this optimization, the SVM model with RBF kernel for ASCADf required 3383 seconds and ASCADv took 42116 seconds for training, which is expected due to substantially larger dataset size. The key recovery was achieved at ~320 traces for both ASCADf and ASCADv. However, the execution was very computationally extensive suggesting this model might not be very efficient for this task. Standard scikit-learn's SVC implementation is often limited in its internal parallelization for training, and SVM in itself is very largely influenced by noisy and irrelevant time points and might not be the most efficient choice for large-scale side-channel analysis. It could be viable in circumstances where noise is less of a factor typically on hardware devices with clearer leakage or in datasets that are de-noised. Future investigations could also explore other high-performance GPU-based SVM libraries.  

\begin{figure}[H]
    \centering
    \begin{minipage}[b]{0.48\textwidth}
        \centering
        \includegraphics[width=\textwidth]{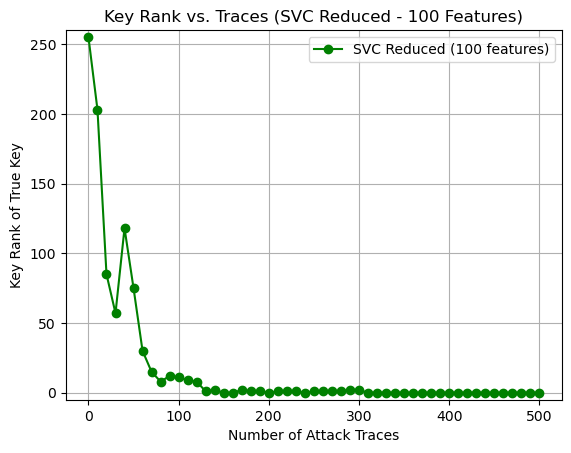}
    \end{minipage}
    \hfill
    \begin{minipage}[b]{0.48\textwidth}
                \centering
                \scalebox{1}[1.22]{\includegraphics[width=\textwidth]{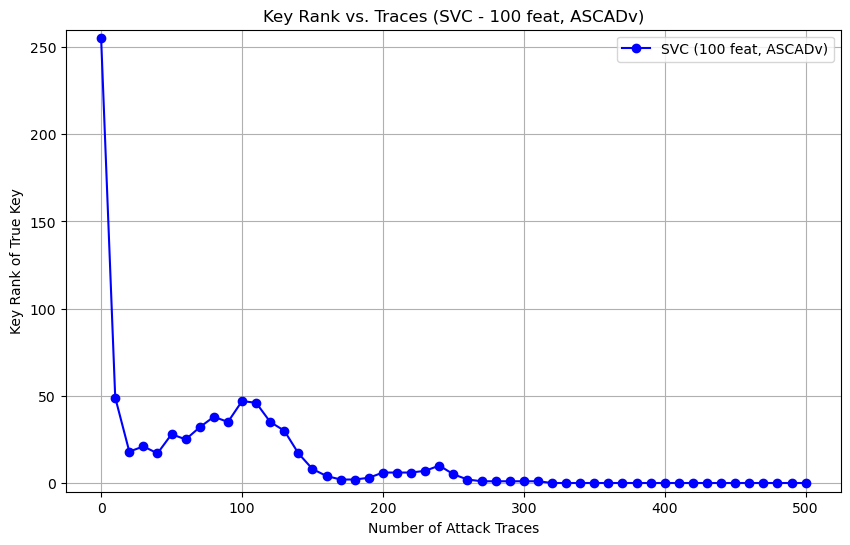}}
        \end{minipage}
    \caption{Key Rank Charts for SVC on Datasets ASCADf(left) and ASCADv(right)}
    \label{fig:rf_key_rank_combined}
\end{figure}

\subsection{ResNet Results}
ResNets turned out to be one of the best-performing models. On the ASCADf dataset, the ResNet performed comparably to the standard CNN, successfully recovering the key after 110 traces. The model demonstrated a robust ability to learn the necessary leakage patterns from the fixed-key training data.
The true value of the ResNet architecture was revealed on the ASCADv dataset. Where the standard CNN had previously failed, the ResNet model succeeded unequivocally. The key recovery succeeded in just 30 traces. This performance of significantly fewer traces than on the `easier' fixed-key dataset was something of a surprise but aligns with Karayalcin et al., whose study also reached similar conclusions and noted that ResNets are particularly suited to datasets with large training data like ASCADv. Although reaching similar conclusions, our ResNet on ASCADf and ASCADv performs significantly better than Karayalcin et al. These results from ResNet confirm that the addition of residual connections is a critical architectural improvement, which enables the network to effectively solve complex, variable-key side-channel challenges.

\begin{figure}[H]
    \centering
    \begin{minipage}[b]{0.48\textwidth}
        \centering
        \scalebox{1}[1.1]{\includegraphics[width=\textwidth]{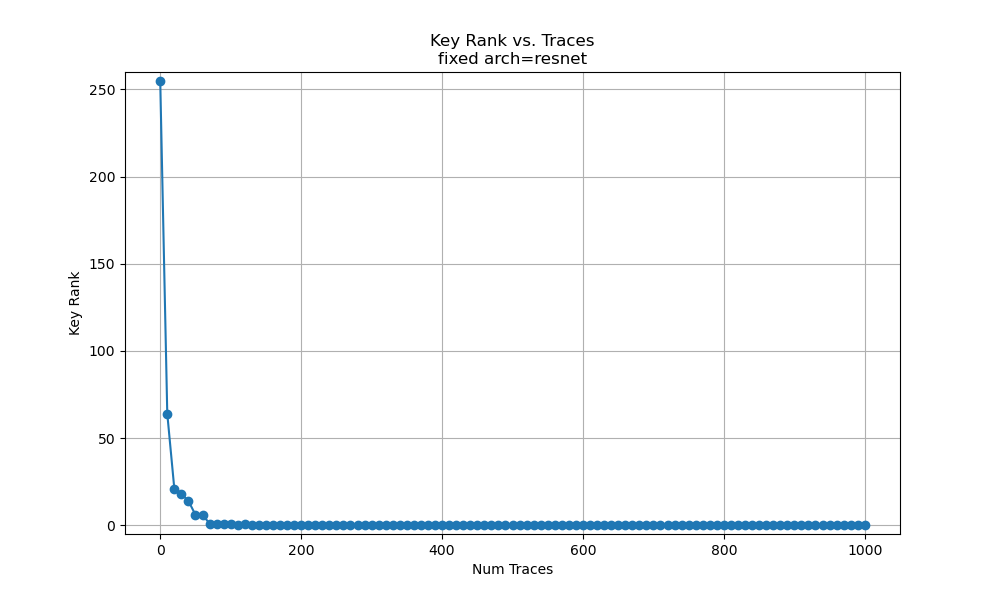}}
    \end{minipage}
    \hfill
    \begin{minipage}[b]{0.48\textwidth}
                \centering
                \scalebox{1}[1.1]{\includegraphics[width=\textwidth]{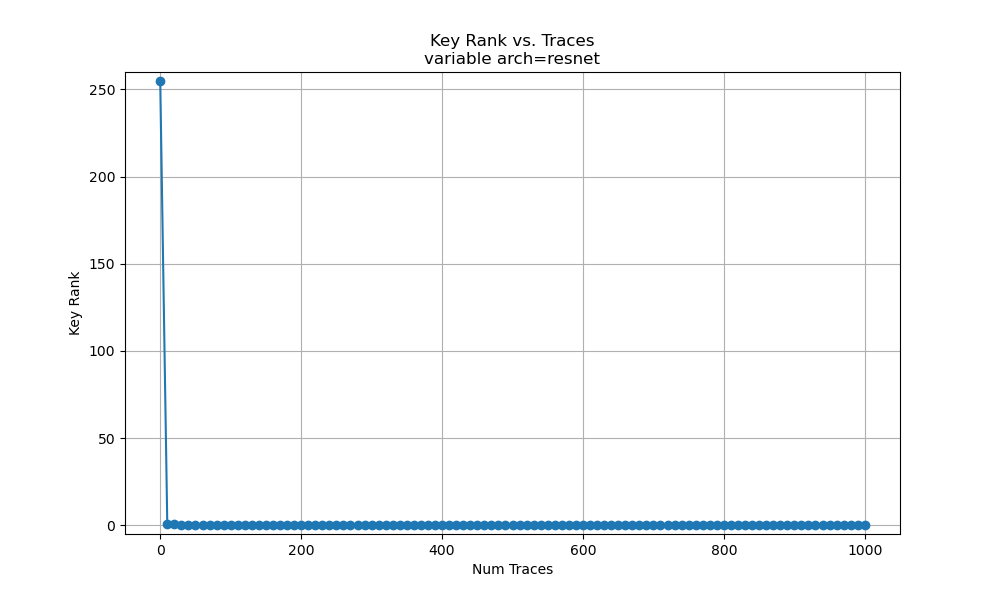}}
        \end{minipage}
    \caption{Key Rank Charts for ResNets on Datasets ASCADf(left) and ASCADv(right)}
    \label{fig:rf_key_rank_combined}
\end{figure}

\subsection{Summary of Results}
\begin{table}[H]
    \centering
    \caption{Summary of Model Performance on ASCADf }
    \label{tab:model_comparison}
    \begin{tabular}{|p{2.5cm}|p{1.6cm}|p{2.2cm}|p{2cm}|p{2.6cm}|}
        \hline
        \textbf{Model} & \textbf{Features} & \textbf{Traces to Rank 0} & \textbf{Training Time (s)\footnote{}} & \textbf{Key Recovery} \\
        \hline
        RF (full) & 700 & $\sim$492 & 9.36 & Success \\
        \hline
        RF (reduced) & 100 & $\sim$200 & 4.60  & Success \\
        \hline
        SVM (reduced) & 100 & $\sim$320 & 3383 & Success \\
        \hline
        CNN & 700 & $\sim$65 & 9322  & Success \\
        \hline
        ResNets & 700 & $\sim$110 & 5623  & Success \\
        \hline
    \end{tabular}
\end{table}
\footnotetext{Training time incorporates the model’s training duration, including the overhead
for enabling class probability estimation necessary for key guess ranking. Training times also may have been affected due to inefficient code for parallelization. }
\begin{table}[H]
    \centering
    \caption{Summary of Model Performance on ASCADv }
    \label{tab:model_comparison}
    \begin{tabular}{|p{2.5cm}|p{1.6cm}|p{2.2cm}|p{2cm}|p{2.6cm}|}
        \hline
        \textbf{Model} & \textbf{Features} & \textbf{Traces to Rank 0} & \textbf{Training Time (s)\footnotemark } & \textbf{Key Recovery} \\
        \hline
        RF (full) & 1400 & $\sim$750 & 68.32 & Success \\
        \hline
        RF (reduced) & 100 & $\sim$470 & 17.73  & Success \\
        \hline
        SVM (reduced) & 100 & $\sim$320 & 42116& Success \\
        \hline
        CNN & 1400 & --- & ---  & Failed \\
        \hline
        ResNets & 1400 & ~30 & 46209  & Success \\
        \hline
    \end{tabular}
\end{table}
    
\footnotetext{
The substantially longer training times on ASCADv is expected due to it having 200,000 profiling traces and 100,000 attack traces compared to 50,000 profiling and 10,000 attack traces for ASCADf. }

\section{Discussion}
The experiments conducted demonstrate successful AES key recovery from EM leakage using Machine Learning on the ASCAD datasets. The ResNets with ASCADv proved to be most efficient followed by standard CNN with ASCADf when it achieved Rank 0 after only 65 traces. ResNets for ASCADf achieved Rank 0 after 110 traces, while RF for ASCADf with feature selection needed around 200, outperforming SVM. The SVM performed poorly despite using the same reduced feature set as RF. This in turn, suggests that ensemble methods like RF might be more robust than kernel methods. A key finding is that despite low classification accuracy, successful key recovery was possible. RF feature selection has significantly improved performance and proved that focusing on relevant features certainly benefits certain models. RF for ASCADv was able to bring down traces needed from almost 750 to 470, which is almost over 40\% better performance.

One possible explanation for the CNN's superior performance compared to feature-selected RF is that CNNs automatically learn hierarchical and nonlinear features directly from the raw EM traces. This allows CNNs to capture subtle temporal dependencies and complex leakage patterns that manual feature selection may overlook. Additionally, the CNN architecture is designed to be robust against noise by leveraging convolutional filters and pooling operations. However, despite these architectural advantages, our results show this generalization did not extend very well to the variable-key challenge, resulting the standard CNN to fail. This is where the ResNet architecture demonstrates its crucial advantage. While implemented as structurally similar to our standard CNN, its use of residual `shortcut' connections improved the gradient flow and enabled more effective feature learning in deeper networks. ResNet succeeded spectacularly on the ASCADv dataset by recovering the key in 30 traces. This result confirms that for complex, variable-key scenarios, the improved training stability offered by the ResNet architecture is essential for achieving a successful attack.\cite{karayalcin2023resolving} Thus, while RF with feature selection focuses on the most individually informative features, it might miss intricate interactions present in the full trace data that the Deep Learning counterparts are designed to exploit.

The collective results from these experiments reveal several key insights into the application of machine learning for side-channel analysis. RF performance proves that focusing on high-importance features can significantly improve attack efficiency. Most significantly, experimentation from the standard CNN to the ResNet gives a clear picture of architectural hierarchy. While simpler architectures can succeed in idealized conditions, successfully attacking more complex and realistic challenges requires more sophisticated models like ResNet that are specifically designed to facilitate deep, robust, and generalizable feature learning. 

\subsection{Comparison with Existing Literature}
The results of this study align with existing literature on ML-based SCA, particularly the effectiveness of CNNs and feature selection techniques. For instance, recent works have shown that CNNs can outperform traditional statistical methods in key recovery tasks, especially when dealing with high-dimensional data like EM traces. The use of RF for feature selection is also consistent with findings that highlight its utility in reducing dimensionality and improving model performance in SCA contexts.
\begin{table}[H]
    \centering
    \caption{Comparison with Existing Literature}
    \label{tab:literature_comparison}
    \begin{tabular}{|p{3.5cm}|p{3.5cm}|p{3.5cm}|}
        
        \hline
        \textbf{Reference} & \textbf{Model} & \textbf{Traces to Rank 0} \\
        \hline
        This Study(Best) & RF (reduced) & $\sim$200  \\
        \hline
        This Study(Best) & CNN & $\sim$65  \\
        \hline
        This Study(Best) & ResNets & $\sim$30  \\
        \hline
        Huang et al.\cite{huang2025deep} & Inception Net & $\sim$30   \\
        \hline
        Rousselot et al. \cite{rousselot2025scoop} & Scoop - CNN & $\sim$73  \\
        \hline
        Zaid et al. \cite{zaid2020methodology} & CNN & $\sim$191 \\
        \hline
        Karayalcin et al. \cite{karayalcin2023resolving} & ResNets & $\sim$47 \\
        \hline
    \end{tabular}
\end{table}

\subsection{Potential Countermeasures}
The successful key recovery demonstrates that even standard AES implementations on common microcontrollers are vulnerable to ML-based EM SCA if no specific countermeasures are in place. Potential countermeasures that are commonly used to mitigate such attacks include:
\begin{itemize}
    \item[-]\textbf{Hardware-level:} Noise generation, power supply randomization, specific chip design to reduce EM leakage, shielding.
    \item[-]\textbf{Software-level:} Masking (splitting sensitive values into shares processed independently), shuffling (randomizing the order of operations), or constant-time implementations.
\end{itemize}
These countermeasures aim to reduce the correlation between the EM emissions and the processed data, making it more difficult for attackers to extract sensitive information. However, they often come with trade-offs in terms of performance, complexity, and cost. 

\subsection{Analysis of CNN failure on ASCADv and ResNet Performance}
A key finding from our results is the stark performance difference of our baseline CNN architecture between fixed-key and variable-key datasets. While the CNN introduced in section 2.4 consistently recovered the key on ASCADf, it failed completely on the more challenging ASCADv dataset. This hints that this architecture, while effective at learning specific patterns, lacks the ability to generalize to the more complex scenario where the secret key is not constant. We attribute this limitation primarily to the challenges of training deeper networks, where issues like the vanishing gradient can prevent the model from learning the more abstract features needed for a variable-key attack.

To investigate this limitation further and test our hypothesis, we implemented a follow-up experiment using a Residual Network (ResNet) architecture. ResNets are specifically designed to overcome the challenges of training deep networks through the use of residual skip connections, which allow gradients to flow more effectively during training.\cite{he2016deep} This results in an architecture that is more capable of learning the abstract and long-range feature dependencies that is required to generalize across the non-stationary signals. 

The results of this follow-up experiment were remarkable yet surprising. The ResNet model proved highly effective in recovering the key on the ASCADv dataset in approximately 20 traces, while it required 110 traces on the simpler ASCADf dataset. The exceptional performance on the variable-key set aligns with findings from Karayalcin et al., who also note that ResNets are particularly well-suited for larger, more complex datasets where their architectural depth can be fully leveraged.  \cite{karayalcin2023resolving}

This relative inefficiency of the more complex model on a simpler task can be attributed to architectural overkill. On the ASCADf dataset, the baseline CNN was already sufficient and highly efficient ($\sim 65$ traces). The additional depth and complexity of the ResNet proved to be less efficient. This is likely due to two factors. First, the skip connections may have created a more complex optimization landscape that hindered rapid convergence when the leakage patterns were relatively simple. Second, a deeper network is incentivized to find abstract feature combinations; on a dataset with straightforward, localized leakage like ASCADf, this can be counterproductive, causing the model to take longer to converge on the simpler, more direct patterns.

This confirms that optimal deep learning architecture for side-channel analysis is very dependent on the target's complexity. For simpler, fixed-key targets, a well-tuned but shallower CNN can be more efficient and effective but for more challenging, real-world scenarios involving variable keys, more advanced architectures like ResNet is optimal to achieve efficient results.

\subsection{Limitations}
The study has its some limitations inherent to its methodology such as it primarily relies on the use of synchronized dataset and the profiling attack methodology, which represents a somewhat best-case scenario for the attacker. While this is very important from a vulnerability and technology standpoint this may not be a efficient practical implementation in terms of real-world application and penetration. Real-world attacks might face significant timing jitter (desynchronization) which can further increase complexity. Adapting these models to get efficient results in more realistic conditions would require specialized preprocessing or few architectural changes. More advanced architectures incorporating attention mechanisms could also be explored, as they can learn to focus on relevant leakage patterns regardless of their exact temporal position. Our profiling attack also assumes known plaintext during the attack phase for key ranking, which may not always be available. The trained models are also specific to the ASCAD dataset and may not generalize well to other datasets or other devices for that matter.

\section{Conclusion}
This work has successfully applied and compared Random Forest, SVM, Convolutional Neural Network, and ResNet models for AES key recovery via electromagnetic side-channel analysis on the public ASCAD dataset. We demonstrated that despite extremely low classification accuracy, the Key Rank metric revealed successful and efficient key byte recovery using a tailored CNN and ResNets and a Random Forest trained on features selected via importance ranking. Our findings confirm that ML techniques can effectively learn subtle leakage patterns, even when models exhibit overfitting by standard validation metrics. Feature reduction using RF importance analysis significantly improved performance, reducing the number of attack traces required for successful key recovery. Deep learning models stood out for their efficiency in learning relevant features directly from raw EM traces. The SVC model, while theoretically powerful, struggled to achieve similar performance, suggesting that ensemble methods like RF may be more robust in this context. The results underscore the practical vulnerability of AES implementations to ML-driven side-channel attacks, highlighting the need for effective countermeasures and robust security evaluation methodologies.

Future work could involve applying these techniques to more challenging scenarios like the desynchronized ASCAD datasets, exploring alternative feature selection methods beyond RF importance, and developing models that can recover multiple key bytes simultaneously to extract the full AES key.\cite{berreby2023investigating} Applying these ML models to datasets from diverse hardware platforms and against AES implementations protected with known countermeasures (e.g., masking) would provide valuable insights into the practical resilience of these defenses. Architecturally, exploring attention mechanisms or residual connections within CNNs could enhance feature learning and potentially improve performance, especially with noisy or desynchronized traces. 

Ultimately, this research highlights the tangible threat posed by ML-driven side-channel attacks, reinforcing the critical need for robust hardware and software countermeasures, alongside rigorous security evaluation methodologies, to protect cryptographic implementations in real-world devices.

\section*{Acknowledgments}
This work was supported by the Jimmy Payne Foundation through the Computing Research Center Grant at the University of Southern Mississippi. The authors acknowledge HPC at The University of Southern Mississippi supported by the National Science Foundation under the Major Research Instrumentation (MRI) program via Grant \#ACI 1626217.

\bibliography{sn-bibliography}

\begin{thebibliography}{10}
\providecommand{\url}[1]{\texttt{#1}}
\providecommand{\urlprefix}{URL }

\bibitem{obaid2025enhancing}
Obaid, Z.M., Ali~Alheeti, K.M.: Enhancing malware detection through electromagnetic side-channel analysis using random forest classifier. Journal of Cybersecurity \& Information Management  15(2) (2025)

\bibitem{berreby2023investigating}
Berreby, Y.E., Sauvage, L.: Investigating efficient deep learning architectures for side-channel attacks on aes. arXiv preprint arXiv:2309.13170  (2023)

\bibitem{kocher1999differential}
Kocher, P., Jaffe, J., Jun, B.: Differential power analysis. In: Advances in Cryptology—CRYPTO’99: 19th Annual International Cryptology Conference Santa Barbara, California, USA, August 15--19, 1999 Proceedings 19. pp. 388--397. Springer (1999)

\bibitem{benadjila2020deep}
Benadjila, R., Prouff, E., Strullu, R., Cagli, E., Dumas, C.: Deep learning for side-channel analysis and introduction to ascad database. Journal of Cryptographic Engineering  10(2),  163--188 (2020)

\bibitem{huang2025deep}
Huang, H., Wu, J., Tang, X., Zhao, S., Liu, Z., Yu, B.: Deep learning-based improved side-channel attacks using data denoising and feature fusion. PloS one  20(4),  e0315340 (2025)

\bibitem{aes_round_commons}
{John Savard}: {AES (Rijndael) Round Function} [png image]. Wikimedia Commons (1999), \url{https://commons.wikimedia.org/wiki/File:AES_(Rijndael)_Round_Function.png}, dedicated to public domain under CC0 1.0 Universal (https://creativecommons.org/publicdomain/zero/1.0/)

\bibitem{mangard2006pinpointing}
Mangard, S., Schramm, K.: Pinpointing the side-channel leakage of masked aes hardware implementations. In: International Workshop on Cryptographic Hardware and Embedded Systems. pp. 76--90. Springer (2006)

\bibitem{he2016deep}
He, K., Zhang, X., Ren, S., Sun, J.: Deep residual learning for image recognition. In: Proceedings of the IEEE conference on computer vision and pattern recognition. pp. 770--778 (2016)

\bibitem{picek2019curse}
Picek, S., Heuser, A., Jovic, A., Bhasin, S., Regazzoni, F.: The curse of class imbalance and conflicting metrics with machine learning for side-channel evaluations. IACR Transactions on Cryptographic Hardware and Embedded Systems pp. 209--237 (2019)

\bibitem{karayalcin2023resolving}
Karayalcin, S., Perin, G., Picek, S.: Resolving the doubts: On the construction and use of resnets for side-channel analysis. Mathematics  11(15),  3265 (2023)

\bibitem{rousselot2025scoop}
Rousselot, N., Heydemann, K., Masure, L., Migairou, V.: Scoop: An optimizer for profiling attacks against higher-order masking. Cryptology ePrint Archive  (2025)

\bibitem{zaid2020methodology}
Zaid, G., Bossuet, L., Habrard, A., Venelli, A.: Methodology for efficient cnn architectures in profiling attacks. IACR Transactions on Cryptographic Hardware and Embedded Systems pp. 1--36 (2020)

\end{thebibliography}
\bibliographystyle{sn-apacite}

\end{document}